\documentclass[mathleft
]{an}
\usepackage{graphicx}
\usepackage{times}
\usepackage{amsmath}
\begin{document}

\def\mstrichi{\begin{picture}(0,0)\put(2,7){\rule{0.6em}{0.3pt}}\end{picture}m}
\def\gammastr{\bar{\gamma}}

\Pagespan{789}{}
\Yearpublication{2006}%
\Yearsubmission{2005}%
\Month{11}%
\Volume{999}%
\Issue{88}%

\title{Metal enrichment of the intracluster medium: SN-driven galactic winds}

\author{Verena Baumgartner\inst{1}
\and  Dieter Breitschwerdt\inst{2}
}
\titlerunning{SN-driven Galactic Winds}
\authorrunning{V. Baumgartner \& D. Breitschwerdt}
\institute{
Institut f\"ur Astronomie, University of Vienna,T\"urkenschanzstr. 17, A-1180 Vienna, Austria
\and
Zentrum f\"ur Astronomie und Astrophysik, Technische Universit\"at Berlin, Hardenbergstr. 36, D-10623 Berlin, Germany
}

\received{}
\accepted{}
\publonline{}


\abstract{%
We investigate the role of supernova (SN)-driven galactic winds in the chemical enrichment
of the intracluster medium (ICM). Such outflows on galactic scales have their origin in huge star forming regions and
expel metal enriched material out of the galaxies into their
surroundings as observed, for example, in the nearby starburst galaxy NGC 253.
As massive stars in OB-associations explode sequentially, shock waves are
driven into the interstellar medium (ISM) of a galaxy and merge, forming a
superbubble (SB). These SBs expand in a direction perpendicular to the disk
plane following the density gradient of the ISM. We use the 2D analytical
approximation by Kompaneets (1960) to model the expansion of SBs in an
exponentially stratified ISM. This is modified in order to describe the sequence
of SN-explosions as a time-dependent process taking into account the main-sequence life-time of the SN-progenitors and using an
initial mass function to get the number of massive
stars per mass interval. The evolution of the bubble in space and time is
calculated analytically, from which the onset of Rayleigh-Taylor instabilities in
the shell can be determined. In its further evolution, the shell will break up and
high-metallicity gas will be ejected into the halo of the galaxy and even into the
ICM. We derive the number of stars needed for blow-out depending on the
scale height and density of the ambient medium, as well as the fraction of alpha-
and iron peak elements contained in the hot gas.
Finally, the amount of metals injected by Milky Way-type galaxies
to the ICM is calculated confirming the importance of this enrichment process.
}

\maketitle

\section{Introduction}
\begin{figure}
\begin{center}
\includegraphics[width=0.45\textwidth]{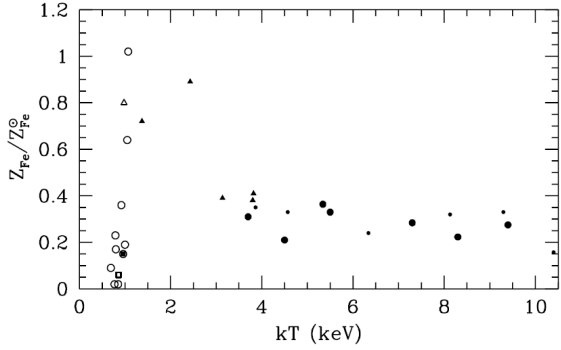}
\caption{\label{Renzini}Iron abundances as a function of ICM temperature for clusters and groups.
Clusters at moderately high redshift $\langle z \rangle \simeq 0.35$ are shown by small filled
circles (\cite{r97}).}
\end{center}
\end{figure}

The space between the galaxies in a cluster is filled with a low-density, high-temperature plasma, the so-called intracluster medium (ICM). It constitutes the largest part of the visible mass of a cluster. X-ray observations reveal that the ICM contains heavy elements like iron, silicon and oxygen instead of existing entirely of primordial gas. On average, metallicities range from 1/3 to 1/2 of the solar abundances (e.g.~\cite{r97}, \cite{f98}) (see Fig.~\ref{Renzini}).
Already in 1975, Larson and Dinerstein discussed the chemical enrichment of the ICM in their theoretical paper and suggested that mass loss from galaxies in a cluster increases the metallicity of the ICM. Shortly after that, the iron emission line was detected in the X-ray spectra of the Perseus cluster (Mitchell 1975). A large number of samples of nearby clusters (e.g.~Mushotzky et al. 1978; \cite{w2000}) has been analyzed until now, as well as clusters out to redshifts of $z\,\sim\,1$ (e.g. Tozzi et al. 2003, Balestra et al. 2007). Only a small evolutionary trend can be seen in the abundances. It is obvious that investigations of the enrichment of the ICM help us to understand the origin of the ICM and the star formation history of the universe.
Since heavy elements are the product of nucleosynthesis inside the cluster galaxies, the question comes up, how they can be transported from the cluster galaxies into the ICM. Possible enrichment processes are galactic winds, ram pressure stripping and jets from active galactic nuclei. Exploring galaxy clusters with X-ray observatories like XMM-Newton results in 2D abundance maps, which show, in general, an inhomogeneous distribution of the metals (Durret et al. 2005). Studying the distribution of metals in the ICM provides constraints on the efficiency of these processes and on the evolution
of clusters and the galaxies within (\cite{sch05}).
We need to know not only how these processes are working but also must calculate the amount of metals that are ejected or stripped in order to learn about the formation and chemical evolution of galaxies.

\section{Galactic outflows}
%
Galactic winds are powerful outflows of galaxies originating from massive bursts of star formation. Thus they are a common phenomenon in starburst galaxies, where the star formation rate is enhanced by a factor of 10 or more. In addition cosmic rays can assist in driving galactic winds, provided that the coupling between high energy particles and gas is sufficiently strong (\cite{B91}, \cite{Ev08}).
Heavy elements are synthesized in massive stars, which end their life as supernovae.
The released energy of successive SNe in star forming regions drives the outflow
and the metal enriched material is expelled into the halo of a galaxy or even into intracluster space.\\
Galactic winds as an enrichment process were proposed by Larson $\&$ Dinerstein (1975) and then by DeYoung (1978). The amount of mass loss depends on properties of a galaxy itself, such as mass and disk scale length, and on the environment.
An important question is, where the transition of metal ejection to metal retention is.
Wiebe et al.~(1999) find that galaxies with more than $5 \cdot 10^{11}$ M$_{\odot}$ do not eject any material at all.
Dwarf galaxies may suffer a complete blow-away due to their shallow potentials, while galaxies with $10^9$ M$_{\odot}$ lose less than 3 $\%$ of the metal enriched gas (\cite{mlf99}).
With outflow speeds of about 200-1500 km/s (\cite{BH07}), the escape velocity from the gravitational potential is easily exceeded in low-mass galaxies, such that the gas is lost from the galaxies.
The other possible fate of the high-metallicity gas is that it cools and falls back onto the galaxy, becoming part of the galactic fountain (\cite{sh76}) and consequently returning the newly synthesized material to the ISM.
Contrary to ram pressure stripping, galactic winds are usually suppressed in the central regions of a cluster due to the high ICM pressure, against which outflows have to work.
%
%

\section{Superbubble dynamics}
Highly energetic interaction processes between the components of the ISM are found where strong explosions like
supernovae occur or stellar winds from massive stars sweep up the ISM.
Massive stars are usually found in associations rather than randomly in the field (\cite{ll03}). They are strongly correlated in space in regions of $\sim$100$\,$pc radius and in time.
The stellar winds from these stars and multiple SN-explosions produce shock waves and carve out large holes filled with hot, tenuous and metal enriched gas. These SBs can blow out of the disk and expel the hot gas into the halo. Moreover, they deliver energy to the ISM and into the halo, and hence provide important information about the heating of the disk and the galactic corona by SNe.\\
The aim of our work was to develop a time-dependent mo\-del for superbubble evolution to find out if galactic winds can be a dominant enrichment process, and to estimate how many metals are released in a wind.

\subsection{Kompaneets' approximation}
\begin{figure}
\begin{center}
\includegraphics[angle=270,width=0.45\textwidth]{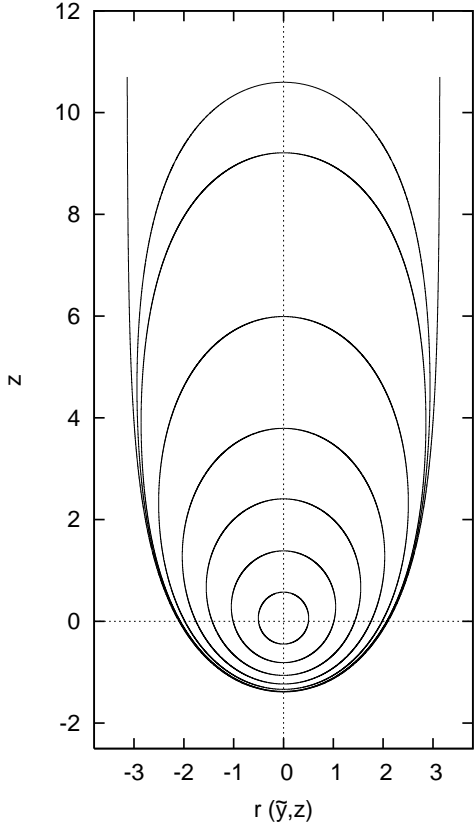}
\caption{\label{SBs}Position of the shock front at $\tilde{y} = 0.5$, $1.0$, $1.4$, $1.7$, $1.9$, $1.98$, $1.99$, and $2.0$.}
\end{center}
\end{figure}

%
At early times in the evolution of a superbubble
the surrounding medium is assumed to appear homogeneous and therefore the surface of the shock front remains spherical.
The similarity solution (\cite{sed46}, \cite{t50}) can be used to describe the expansion of the bubble in these early stages as long as the radius is less than the ambient density scale height.
Due to the negative density gradient the bubble becomes elongated after reaching one scale height above the galactic plane (Fig.~\ref{SBs}).
We use the 2D analytical approximation found by Kompaneets (1960), which -- in its original form -- describes the propagation of an adiabatic shock wave generated by a point explosion in an exponentially stratified medium. The density variation is given by
$
\rho(z) = \rho_0 \,\exp (-z/H).
$
Here, $z$ is the height above the galactic plane, $\rho_0$ is the density at $z=0$ and $H$ is the scale height of the interstellar
gas.
In our work, we use Kompaneets' approximation (hereafter KA) to determine the shape and the time evolution of a superbubble analytically, which provides a proper insight into the physics of bubble evolution.\\
The axially symmetric problem is described in cylindrical coordinates $(r, \, z)$. In order to use the
KA for the investigation of SB evolution the following assumptions have to be made (see also \cite{s85}; \cite{bks95}; \cite{b99}):
(i) a strong shock is propagating into the undisturbed ISM. This means that the pressure in front of the shock, i.e. of the ambient gas is negligible; (ii) the pressure of the shocked gas is spatially uniform and (iii) almost all the swept-up gas behind the shock front is located in a thin shell.
In the analysis of the evolution of the shock surface, the Rankine-Hugoniot jump conditions have to be solved at every point of the shock surface (\cite{bks95}; \cite{b99}). The solution is performed using a transformed time variable (in units of a length), which represents the evolutionary status of a bubble
\begin{equation}
y =  \int_{0}^{t} \sqrt{\frac{\gamma^2 - 1}{2}\frac{E_{\rm{th}}}{\rho_0 \cdot \Omega}} \, dt
\label{y}
\end{equation}
with $E_{\rm{th}}$ being the thermal energy of the shocked gas and $\Omega$ being the volume confined by the shock front. A ratio of specific heats $\gamma=5/3$ is used.
The result is the half-width extension of the bubble parallel to the galactic plane (see Fig.~\ref{SBs})
\begin{equation}
r(y, z)=2H\arccos \left[ \frac{1}{2} e^{z/2H} \left(1\!-\!\frac{y^2}{4H^2} + e^{-z/H} \right) \right].
\end{equation}
Top and bottom of the bubble are given by
\begin{equation}
z_u(\tilde{y}) = -2H \cdot \ln\ (1-\tilde{y}/2)
\label{zu}
\end{equation}
and
\begin{equation}
z_d(\tilde{y}) = -2H \cdot \ln\ (1+\tilde{y}/2),
\end{equation}
respectively, with $\tilde{y} = y/H$.
%
%

\subsection{Superbubbles in stratified disks}
In order to derive analytical expressions for time, velocity and acceleration of the superbubble we approximate the volume of the bubble by an ellipsoid\footnote{\label{foot:1} Also Maciejewski and Cox (1999) approximate the shape of the shock front by a prolate ellipsoid.}, with a semimajor axis
$a(\tilde{y}) = (z_u(\tilde{y}) - z_d(\tilde{y}))/2$
and a semiminor axis
$r_{\rm{max}}(\tilde{y})=2H \, \arcsin \, (\tilde{y}/2)$.
Thus, we get the bubble volume $V(\tilde{y}) = \frac{4 \pi}{3} \, a(\tilde{y}) \, r^2_{\rm{max}}$ as a function of the time variable.
The deviation from the numerical integration of the volume amounts to only $\sim$ 2 $\%$ at very late stages of evolution ($\tilde{y}$ = 1.9).\\
In our description of superbubble evolution we want the initial point-like energy deposition in the KA to be replaced by a time-dependent energy input rate.
We calculate the energy thermalized at the inner shock following Weaver et al. (1977), but using the time-dependent energy input rate
$L_{\rm{SB}}(t)= L_{\rm{IMF}} \cdot t^{\delta}$ instead of a constant wind.
In such a model, masses and numbers of stars are given by an Initial Mass Function (e.g.~\cite{f06})
\begin{equation}
\frac{dN(m)}{dm} = N_0 \cdot m^{\bar{\gamma}}.
\end{equation}
With $\Gamma = \bar{\gamma} + 1$ and by fixing a stellar mass range, the integration of the equation above results in the total number of stars between a lower mass limit $m_l$ and an upper mass limit $m_u$
\begin{equation}
N_{\rm{OB}} = \frac{N_0}{\Gamma} \cdot  \left. m^{\Gamma} \right|_{m_l}^{m_u}
\end{equation}
with masses in solar units.
The normalization factor $N_0$ is given by the number of stars belonging to the OB-associa\-tion. We obtain it by fixing
the number of stars in the last mass bin ($m_u-1, m_u$) to be $N_{\rm{OB}} = 1$.
%
%
\begin{figure}
\includegraphics[angle=270,width=0.45\textwidth]{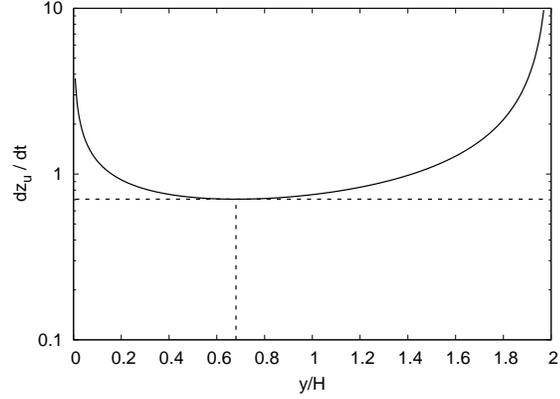}
\caption{\label{velocity}Dimensionless velocity (solid line) at the top of the bubble $z_u$ as a function of $\tilde{y}$. The minimum of the velocity is reached where $\tilde{y}_{\rm{acc}} = 0.68$ (dashed line). }
\end{figure}

\noindent
Treating the number of stars between $(m, m + dm)$
as a function of mass and expressing the stellar mass by its main sequence lifetime $(t, t - dt)$ in order
to obtain a time-sequen\-ce of exploding stars (\cite{bb02}) results in
\begin{equation}
L_{\rm{SB}}(t) = E_{\rm{SN}}\frac{dN(m)}{dm} \cdot \left(-\frac{dm}{dt} \right).
\end{equation}
The main sequence life time for massive stars as a function of mass (in units of seconds) can be described by a power-law
$t (m) = \kappa m^{-\eta}$.
Values taken from Fuchs et al. (2006) are $\kappa = 1.6 \cdot 10^{8} \rm{yr} $ and $\eta = 0.932$.
Rearranging leads to the mass of a star expressed by its main sequence lifetime
$
m(t) =  (t/\kappa )^{-1/\eta}
$.
We get
\begin{equation}
L_{\rm{SB}}(t)=\frac{E_{\rm{SN}} \cdot N_0}{\eta \cdot \kappa} \left ( \frac{t}{\kappa} \right )^{-( \frac{\Gamma}{\eta}+1)}.
\end{equation}
Since $L_{\rm{SB}}(t)$ is $\propto t^{\delta}$, we have
$
\delta = - ( \frac{\Gamma}{\eta}+1 )
$.
Finally, the result for the thermal energy as a function of time is
\begin{equation}
E_{\rm{IMF}}(t)= \frac{5}{7\delta + 11} \, L_{\rm{IMF}} \cdot t^{\delta+1}.
\end{equation}
Checking for a constant energy input rate with $\delta=0$ corresponds to the Weaver et al. (1977) solution with a constant factor of 5/11.
We use the energy input rate, the density and the scale height of the interstellar medium to define a
time scale in units of seconds $t_{0} =(\rho_0 \cdot H^5/L_{\rm{IMF}})^{(1/(\delta +3))}$.
By rearranging equ.~\ref{y} we obtain the time as a function of $\tilde{y}$
\begin{equation}
t(\tilde{y}) = t_{0} \cdot \left( \frac{(\delta + 3)^2 (7 \delta + 11)}{20\cdot \beta^2} \right) ^{\alpha}
\left( \int_0^{\tilde{y}} \sqrt{\tilde{V}(\tilde{y}')}\,d\tilde{y}' \right)^{2\alpha}
\label{time}
\end{equation}
with $\alpha= 1/(\delta + 3)$ and $\beta = \sqrt{\frac{\gamma^2 - 1}{2}}$.
By expressing $N_0$ as a function of $N_{\rm{OB}}$ and the integral of the square root of the volume by a simple power law, we derive the time in typical units of the ISM
\begin{equation}
\begin{split}
t_{\rm{ap}}(\tilde{y}) &= 7.85 \cdot \left( \frac{n_0}{1\, \rm{cm}^{-3}} \right) ^{\alpha} \left( \frac{H}{100 \, \rm{pc}}
\right) ^{5\alpha}
\left( N_{\rm{OB}}\right)^{-0.87\alpha} \\
&\cdot (\tilde{y}^{2.919})^{2\alpha} \, \rm{Myr}.
\end{split}
\end{equation}
The errors for this approximation are below 2$\,\%$ ($\tilde{y}=1.9$, $\sim \,$100 stars, Lockman layer).
%
%
%
\begin{figure}
\begin{center}
\includegraphics[angle=270,width=0.45\textwidth]{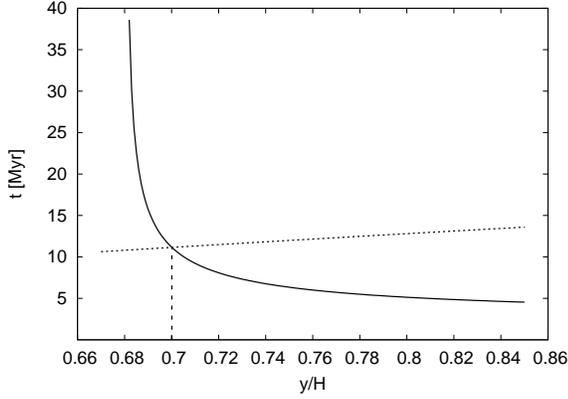}
\caption{\label{tcross}Instability time scale (solid line) and dynamical time scale (dotted line) for n $=$ 0.1~cm$^{-3}$, H $=$ 500~pc, N$_{\rm{OB}}$ $=$ 20 as a function of the time
variable $\tilde{y}$. Acceleration starts at $\tilde{y}_{\rm{acc}} = 0.68$, the curves cross each other at $\tilde{y}_{\rm{rti}} = 0.70$ (indicated by the dashed line).}
\end{center}
\end{figure}

\noindent
Velocity $\dot{z}_u(\tilde{y})$ and acceleration $\ddot{z}_u(\tilde{y})$ of the top of the bubble are derived by calculating the first and second time deri\-vative, respectively, of the coordinate $z_u$ (equ.~\ref{zu}).
Using different formulas
for the main sequence life time or also varying the slope of the IMF may change the value of $\tilde{y}_{\rm{acc}}$ and hence the point in time where the acceleration sets in.
For an IMF with slope $\Gamma = -1.35$ (\cite{s55}) we see that the velocity has its minimum at $\tilde{y}_{\rm{acc}} = 0.68$ (Fig.~\ref{velocity}).
By assuming that the shock has to stay strong all the time we introduce our blow-out condition: the velocity of the outer shock needs to have a Mach number of at least $M \ge 5$.
We find that blow-out into the Lockman-layer (\cite{lock84}) of a Milky Way-like galaxy (n = 0.1$\,$cm$^{-3}$, H = 500$\,$pc, and T = 6000~K) requires an association with about 20 OB-stars, which, incidentally corresponds to the initial stellar content of the Local Bubble.\\
The acceleration leads to the development of Rayleigh-Tay\-lor instabilities (RTIs) at the contact discontinuity, since a dense shell accelerated by the tenuous gas inside the bubble constitutes an unstable hydrodynamic configuration.
RTIs grow exponentially on a characteristic timescale $t_{\rm{rti}}$, which is calculated for the top of the bubble (in units of seconds)
\begin{equation}
t_{\rm{rti},z_u}(\tilde{y})= \sqrt{ \frac{\Delta d_e(\tilde{y})}{2 \pi \ddot{z}_u(\tilde{y})} \cdot \frac{4 \rho_0 e^{-\tilde{z}_u(\tilde{y})} + \rho_{\rm{in}}(\tilde{y})} {4 \rho_0 e^{-\tilde{z}_u(\tilde{y})} - \rho_{\rm{in}} (\tilde{y})} }.
\end{equation}
The wavelength of the perturbation is comparable to the shell thickness $\Delta d_e$ and $\rho_{\rm{in}}$ means the density of the hot shocked gas.
%
%
%
\begin{figure}
\begin{center}
\includegraphics[angle=270,width=0.45\textwidth]{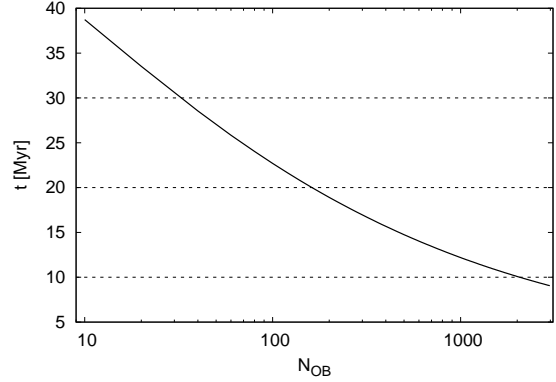}
\caption{\label{fragment}
Fragmentation time scale at the top of the bubble as a function of the total number of stars for associations with 10--3000 OB-stars, again for ISM-values of the Lockman layer. Dashed lines, designating certain time intervals, are used for illustrative purposes.}
\end{center}
\end{figure}

\noindent
If the timescale of an instability to grow is smaller than the dynamical time-scale $t_{\rm{dyn}}(\tilde{y}) = a(\tilde{y})/\dot{z}_u(\tilde{y})$ at any point in time, the shell starts to break up at this moment.
When comparing these two timescales, the value of $\tilde{y}$ can be found, where $t_{\rm{rti},z_u} \leq t_{\rm{dyn}}$.
The break up of the shell starts for all bubble configurations shortly after acceleration sets in, which happens at $\tilde{y}_{\rm{rti}} \cong 0.7$ for rich and poor associations. Fig.~\ref{tcross} shows the behavior of these two curves for an association with 20 OB-stars.
In our calculations it is assumed that the exponentially growing instability is fully developed after approximately $3 \cdot t_{\rm{rti}}$, which we call fragmentation time scale (Fig.~\ref{fragment}). At this point, the instability has reached the fully non-linear regime. The shell will break-up and the hot material inside the bubble can escape and
enrich the surrounding medium with heavy elements. We have to take into account that fragmentation has to occur in a reasonable time scale, before or not long after all SNe have exploded and where shear motions or turbulence in the ISM have no great influence on the shell structure. If we define such a time interval to be 30 Myr after the acceleration at top of the bubble had started, an association needs to have at least 32 OB-stars. In that case, the shell is fragmented at $z_u$ about 41 Myr after the first SN-explosion happened.

\section{ICM metallicities}
\begin{figure}
\begin{center}
\includegraphics[angle=270,width=0.45\textwidth]{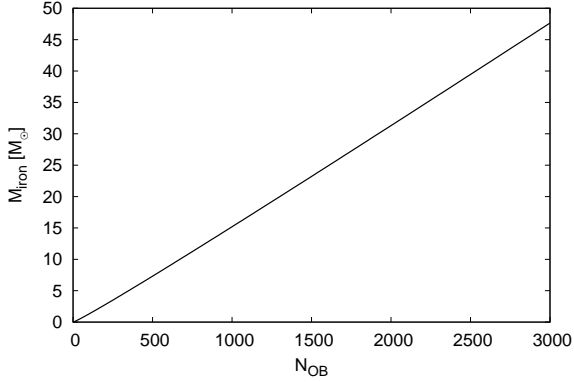}
\caption{\label{iron}Iron mass ejected by SNe in OB-associations. The mass is shown as a function of the total number of OB-stars in the
star cluster using an IMF with $\Gamma=-1.35$.}
\end{center}
\end{figure}

\begin{figure}
\begin{center}
\includegraphics[angle=270,width=0.45\textwidth]{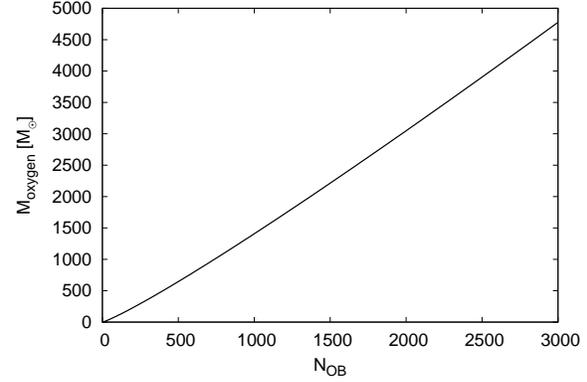}
\caption{\label{oxygen}Same as Fig.~\ref{iron}, but for oxygen.}
\end{center}
\end{figure}

In this section, we calculate the amount of metals ejected in a wind, in order to compare it to the metal content of a galaxy cluster.
Yields for the elements iron (${}^{56} \rm{Fe}$) and oxygen (${}^{16} \rm{O}$) from Woosley $\&$ Weaver (1995) are adopted to a stellar mass range between 8 and 120$\,M_{\odot}$. For the iron mass produced by a massive star with mass $m$ we find a power law (given in solar masses)
\begin{equation}
M_{\rm{Fe}} (m)= 0.02 \cdot m^{0.27} - 0.03.
\end{equation}
A similar result is obtained for oxygen
\begin{equation}
M_{\rm{O}} (m)= 0.049 \cdot m^{1.162}.
\end{equation}
Integration over the whole mass interval of an OB-associa\-tion yields the total mass of iron and oxygen, respectively, contained in the hot interior of a SB. Fig.~\ref{iron} shows the iron mass ejected by SNe as a function of the total number of OB-stars, Fig.~\ref{oxygen} shows the relation for oxygen.\\
Now, that we know the mass of iron and oxygen for SBs with different richness, the next step is to consider the total fraction of metals expelled by SNe in a whole galaxy over its evolution.
%

\noindent
Assuming a SFR of 5 $M_{\odot}$/yr gives -- over an interval of 10$^{10} \,$yr -- about $5 \cdot 10^{10} \,M_{\odot}$ of material which is used to form stars with masses between 0.08 and 120$\, M_{\odot}$. Additionally, about 10$^{11}\,M_{\odot}$ are supplied by 5 strong starburst episodes (200$\,M_{\odot}$/yr) where each episode lasts 10$^8\,$yr. Altogether, there are $1.5\cdot 10^{11}\,M_{\odot}$ representing a Milky-Way like galaxy. The fraction of mass found in SN-progenitors is about 13$\, \%$.
The available mass of $M_{\rm{OB}}=1.9\cdot10^{10} M_{\odot}$ for the formation of high-mass stars has to be distributed among galactic OB-associations.
Observations show that the number of OB-associations in spiral galaxies follows an Initial Cluster Mass Function (ICMF) with a slope $\Omega = -0.75$ (\cite{do08}).
The number of associations as a function of their total mass can be calculated now.
High mass star formation is assumed to happen only in these associations.
The total mass $M_{\rm{OB}}$ of massive stars is distributed among associations that do not
have more than 3000 OB-stars or a mass $M_u = 6.4 \cdot 10^4 \, M_{\odot}$ in form of OB-stars.
The least massive association has two OB-stars or $M_l \sim 20 \, M_{\odot}$.\\
First, a normalization constant has to be calculated
\begin{equation}
N_{0,\rm{OB}}=\frac{M_{\rm{OB}} \cdot (\Omega + 1)}{M_u^{\Omega+1} - M_l^{\Omega+1}}.
\label{531}
\end{equation}
Then the total number of OB-associations can be derived in this mass interval by integration over the ICMF
\begin{equation}
N_{\rm{tot}, \rm{OB}}(M)= \frac{N_{0,\rm{OB}}}{\Omega} \cdot \left(M_2^{\Omega} - M_1^{\Omega} \right).
\label{532}
\end{equation}
%

\noindent
For $M_1 = M_l$ and $M_2 = M_u$, we get a total number of OB-associations of $ \sim 4.8\cdot10^7$. Since only SBs with at least 32 OB-stars or $M_1=400 \, M_{\odot}$ are taken into account (derived from the fragmentation timescale), the number of OB-associations that may enrich the surrounding intergalactic medium is $ \sim 5\cdot10^6$.
If we sum up the number of OB-associations with a certain mass multiplied with the iron or oxygen mass per cluster, we get the amount of heavy elements that is lost to the galactic neighborhood. We find that $1.46 \cdot 10^7 \, M_{\odot}$ of iron and $1.3 \cdot 10^9 \, M_{\odot}$ of oxygen produced by massive stars are expelled by a Milky Way-type galaxy over $10^{10} \,$yr.\\
However, a significant contribution to the blown-out metals comes from the entrainment of ISM, which has been polluted by generations of previous star clusters. 
Thus we need to calculate the metal amount in the swept-up shell of the SB as well. 
We estimate the abundances in the ISM and thus in the shell to have, on average, about half of the solar metallicity over the entire evolution of the Galaxy (e.g.~\cite{i04}).
We get the mass of the shell of the swept-up ISM by integrating the ellipsoidal shape over the density gradient from $z_d$ to $z_u$. At the moment of onset of RTIs, there are typically $\sim 5 \cdot 10^5 \, M_{\odot}$ of gas in the shell.
We derive from geometrical arguments that the mass of the shell fragments that are lost in the outflow, i.e. the cap of the bubble, amounts to about 10$\,\%$ of the total shell mass.
Using abundances by Grevesse, Asplund $\&$ Sauval (2007), we derive that 30$\, M_{\odot}$ of iron and 135$\, M_{\odot}$ of oxygen per SB are injected to the ICM that way.
Altogether, the shell fragments of one galaxy yield an iron mass of about $1.5 \cdot 10^8 \, M_{\odot}$ and an oxygen mass of about
$6.75 \cdot 10^8 \, M_{\odot}$.\\
Our intention is to compare these results to a cluster of 100 Milky Way-like galaxies. Such a cluster
has a total visible mass of about $1.2 \cdot 10^{14} \,M_{\odot}$.
With 1/3 solar metallicity, there are $\sim 4.4 \cdot 10^{10} \,M_{\odot, \rm{Fe}}$ in a galaxy cluster.
Oxygen abundances are 2/3 of the solar abundances, thus the oxygen mass in the cluster is found to be $\sim 4.1 \cdot 10^{11}\,M_{\odot, \rm{O}}$.
Metals in the shell and in the ejecta yield, finally, for 100 galaxies $1.65 \cdot 10^{10} \,M_{\odot, \rm{Fe}}$ and $1.98 \cdot 10^{11}\,M_{\odot, \rm{O}}$. Comparison of our results to the estimates derived from observations, shows that more than 1/3 of the iron and almost 50$\, \%$ of the oxygen mass in the ICM can be explained by galactic outflows. In both cases, metals in the fragmenting shell as well as those ejected by SNe contribute to the enrichment of the intracluster gas.

\section{Closing remarks}
\begin{enumerate}
\item
Although the KA does not take into account inertia of the thin shell, cooling of the shocked gas and gravity, a good agreement between the bubble geometry derived from this model and that of observed SBs usually is found (e.g.~\cite{g74}, \cite{b99}).
The approximation seems to be a good choice to describe SB evolution in a simple, but realistic way, but instead of a pure exponentially decreasing density law one symmetric to the midplane should be included.\\
\item
Comparison to other IMFs (e.g. $\Gamma = -1.15$ and $\Gamma = -1.7$) reveals that a similar number of stars is needed for blow-out into the Lockman layer (18 and 26 stars, respectively).\\
\item
The calculations of the ICM metallicities, although ba\-sed on a rough estimate only, show convincingly that SN-driven galactic winds are a relevant source of metal-enriched material and play an important role in the enrichment of the ICM.
\end{enumerate}
\acknowledgements
VB is recipient of a DOC-fForte Fellowship of the Austrian Academy of Sciences at the University of Vienna.



\end{document}